# MEASURING SIN²2Θ₁₃ WITH THE DAYA BAY NUCLEAR REACTORS

Yi-Fang Wang

*For the Daya Bay collaboration[i]*

*Institute of High Energy physics, Beijing, 100049, P.R. China*
*[*]E-mail: yfwang@ihep.ac.cn*

Angle $\theta_{13}$ is one of the two unknown neutrino mixing parameters to be determined. Its value may determine the future trend of the neutrino physics. We propose to measure $\sin^2 2\theta_{13}$ with a sensitivity better than 0.01 (90% C.L) at the Daya Bay reactor power plant.

## 1. Introduction

Since the observation of neutrino oscillations in the SuperK, SNO and KamLAND experiments, four out of six neutrino mixing parameters have been measured. The remaining unknown mixing parameters are the mixing angle $\theta_{13}$ and the CP phase $\delta$. With nuclear reactors, $\theta_{13}$ can be unambiguously measured, and a determination with sensitivity of 0.01 in $\sin^2 2\theta_{13}$ is feasible[1].

The Daya Bay experiment, located in Shenzhen, P.R. China, about 55 km away from Hong Kong, possesses two distinctive features: 1) very high antineutrino flux; 2) mountains to suppress cosmic-ray-induced backgrounds in a deep underground laboratory. The Daya Bay nuclear power complex has currently four cores in two groups, and one more group with two cores to be online in 2011. Each core has a thermal power of 2.9 GW, resulting a total of 17.4 GW thermal power.

The experimental layout is shown in Fig.1. There is a far detector at the most sensitive location to $\sin^2 2\theta_{13}$ of about 2km from the core, and there are two near detectors close to the respective reactor clusters to monitor the anti-neutrinos emitted from the cores. Such a configuration can cancel most of the reactor related systematic errors since the measurement is relative [2]. The optimal configuration of the experiment and the associated baseline is obtained with a detailed $\chi^2$ minimization for the best sensitivity to $\sin^2 2\theta_{13}$ taking into account the mountain profile, estimated backgrounds, detector systematic errors and the residual reactor-related errors. The overburden of the underground lab accessed via a horizontal tunnel is about 100m for the near sites and 350m for the far site.

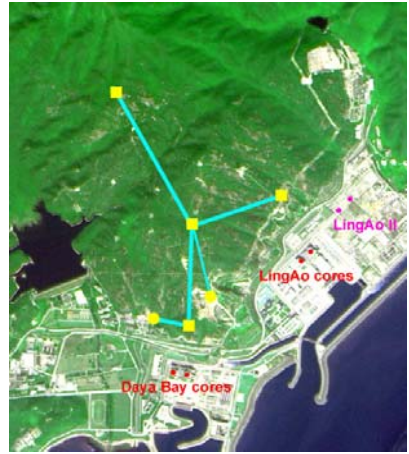

Fig. 1. The Layout of the Daya Bay experiment

## 2. Detector design

To determine $\sin^2 2\theta_{13}$ to the level of 0.01 or better, the statistical and systematic uncertainties have to be controlled to below 0.5%, an improvement of about an order of the magnitude over the past experiments.

Besides using the near-far configuration to cancel the reactor-related systematic errors, the detector-related errors can be reduced by the following means: 1) Employ multiple identical modules at the near and far sites to cross check between modules at each location and reduce detector-related uncorrelated errors. Consideration of cost, calibration of detectors, and statistics results in a design of two modules for each near site and four modules for the far site; 2) Employ a three-zone detector design with zones partitioned by transparent acrylic vessels and well defined target volume. The inner most zone is filled with Gd-loaded liquid scintillator as the antineutrino target. The middle layer (γ-catcher) is filled with normal scintillator to capture any leakage of energy from the target, and the outer most layer (buffer) filled with transparent non-scintillating liquid to shield γ-ray backgrounds from PMT glass and other environmental materials.

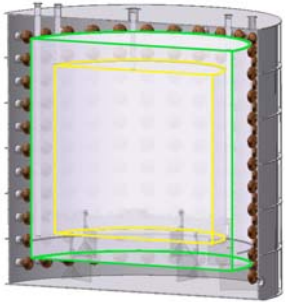

Fig. 2 The three-zone detector module.

As shown in Fig.2, the detector is a cylindrical module with optical reflectors at the top and bottom to improve the uniformity of light response. Based on detailed Monte Carlo simulation, the target volume is determined to be 3.2 m high, 3.2 m in diameter and a total mass of 20 t. The γ-catcher layer is 45-cm-thick, and the buffer layer is also 45-cm-thick. The whole module has a total mass of about 100 t.

Sufficient overburden and shielding are required to reduce γ-rays and energetic cosmic-ray-induced neutrons in the surrounding rock from entering the antineutrino detector modules. Detailed Monte Carlo simulation shows that a water shield with a thickness of more than 2 m can satisfy our requirements. The shield is also a Cherenkov detector. Combining with other kinds of detectors a muon tagging efficiency better than 99.5% with an uncertainty less than 0.25% can be achieved. The baseline design of the muon system is shown in Fig.3. The antineutrino detector modules are immersed in a water pool equipped by about 300 PMTs. Outside of it are water tanks made of reflective PVC sheets with transverse dimensions of 1m × 1m equipped by 4 PMTs at each end. A 13-m-long prototype showed very promising performance of such a detector [3]. At the top of the water pool, three layers of RPC, each with an efficiency of about (90-95)%, are used.

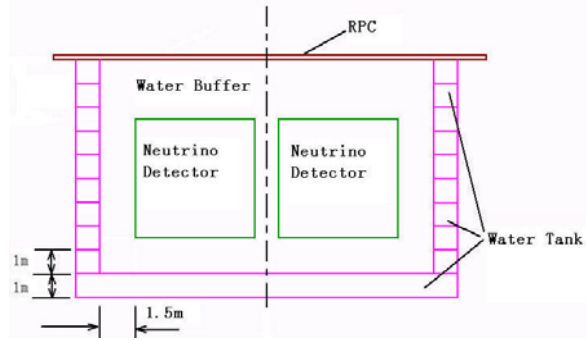

Fig. 3. The veto detector arrangement.

## 3. Backgrounds, systematic errors and the sensitivity

Background control is the key to the success of such a high-precision experiment. There are three important types of background: 1) Radio-activities from the near-by rock, PMT glass and other materials used in the detectors; 2) Cosmic-ray-induced fast neutrons; 3) Cosmic-ray-induced isotopes such $^8$He and $^9$Li. Based on the layout of the apparatus shown in Fig.1 and 2. The estimated amount of background is shown in Table 1.

Table 1. Estimated backgrounds for each site.

|  | DYB | LA | Far |
|---|---|---|---|
| Baseline (m) | 360 | 500 | 1600/1900 |
| Overburden (m) | 98 | 112 | 350 |
| Signal rate (1/day) | 930 | 760 | 90 |
| Natural BK (Hz) | <50 | <50 | <50 |
| Accidental Bk/signal (%) | <0.2 | <0.2 | <0.1 |
| Neutron BK/signal (%) | 0.1 | 0.1 | 0.1 |
| ($^8$He+$^9$Li)/signal (%) | 0.3 | 0.2 | 0.2 |

The difference in the energy spectra of the observed antineutrino signal with and without the oscillation for a $\sin^2 2\theta_{13}$ value of 0.01, along with the estimated background spectra, is shown in Fig. 4.

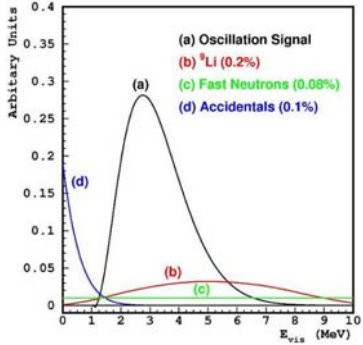

Fig. 4. Energy spectra of neutrino oscillation signals in comparison with estimated backgrounds. The signal is the difference of neutrino energy spectrum with and without oscillation if $\sin^2 2\theta_{13}=0.01$.

Systematic errors of the Daya Bay experiment have to be controlled to the level well below 1%. Based on a detailed analysis of the previous experiments [4] and Monte Carlo simulations, systematic errors are estimated and are listed in Table 2.

The sensitivity of the experiment, based on the configuration shown in Fig.1, is calculated with a $\chi^2$ function that takes into account the expected numbers of antineutrino events from all reactors, systematic errors and backgrounds discussed above along with their estimated uncertainties and correlations. The $\chi^2$ function is constructed as the following:

$$\chi^2 = \min_{\alpha's} \sum_{i=1}^{Nbin} \sum_{A=1,3} \frac{\left[M_i^A - T_i^A(1+\alpha_D+\alpha_c+\alpha_d^A+c_i+\sum_r \frac{T_i^{rA}}{T_i^A}\alpha_r) - b^A B_i^A\right]^2}{T_i^A + T_i^{A2}\sigma_b^2 + B_i^A}$$
$$+ \frac{\alpha_D^2}{\sigma_D^2} + \frac{\alpha_c^2}{\sigma_c^2} + \sum_r \frac{\alpha_r^2}{\sigma_r^2} + \sum_{i=1}^{Nbin} \frac{c_i^2}{\sigma_{shape}^2} + \sum_{A=1,3}\left(\frac{\alpha_d^{A2}}{\sigma_d^2} + \frac{b^{A2}}{\sigma_B^2}\right)$$

where $\sigma_c$ ($\sigma_r$) is the reactor-related correlated (uncorrelated) uncertainty of about 2%; $\sigma_D$ ($\sigma_d$) the detector-related correlated (uncorrelated) uncertainty of about 2% (0.38%); $\sigma_{shp}$ the antineutrino spectral shape uncertainty of about 2%; $\sigma_B^A$ the background-related error as described in Table 1 and $\sigma_b$ the bin-to-bin uncertainty. Fig. 5 shows the sensitivity contours at 90% C.L. for 3 years of data taking from a global $\chi^2$ analysis.

Table 2. Estimated systematic errors in percent per module for the Daya Bay experiment (relative), in comparison with that of the Chooz experiment (absolute)[4]. Baseline values are achievable through proven method while the goals need additional efforts. Those with arrows indicate the case with swapping of the near and the far detector modules.

| Source of uncertainty | | Chooz (Absolute) | Daya Bay(relative) | |
|---|---|---|---|---|
| | | | Baseline | Goal |
| Reactor power | | 0.7 | 0.13 | 0.13 |
| ν spectra | | 2.0 | | |
| ν cross section | | 0.3 | 0 | 0 |
| No. of protons | H/C ratio | 0.8 | 0.2 | 0.1 →0 |
| | Mass | - | 0.2 | 0.02 →0 |
| Eff. | Energy | 0.89 | 0.2 | 0.1 |
| | position | 0.32 | - | - |
| | Time | 0.4 | 0.1 | 0.03 |
| | P/Gd ratio | 1.0 | 0.1 | 0.1 →0 |
| | n multi. | 0.5 | 0.05 | 0.05 |
| | trigger | 0 | 0.01 | 0.01 |
| | Livetime | 0 | <0.01 | <0.01 |
| Back-grounds | correlated | 0.3 | 0.2 | 0.2 |
| | uncorrelated | 0.3 | 0.02 | 0.02 |

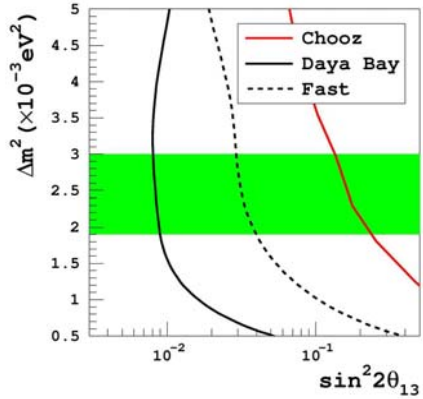

Fig. 5. Expected $\sin^2 2\theta_{13}$ sensitivity at 90% C.L. with 3 years of data.

## 4. Status of the project

The Daya Bay experiment was initiated in 2003. Since then through detailed Monte Carlo simulation and analysis of the previous experiments the baseline layout of the experiment is established.

Our current R&D efforts include: 1) site geotechnical survey including detailed topographic maps, subsurface geophysical survey, and bore drilling that leads to the completion of a conceptual design of the tunnel and experimental halls. Detailed engineering design will start soon; 2) A prototype detector module built at IHEP with a height of 2m and a diameter of 2m, covered with a muon veto detector system. A total of 0.6 t of liquid scintillator contained in an acrylic cylinder is viewed by 45 8" PMTs. Critical design choices, parameters and detector performance are tested and verified. Some results are shown in Fig.6; 3) New type of Gd-loaded liquid scintillator based on Linear Alkylbenzene (LAB) with long attenuation length (~10m) and excellent long-term stability is developed at IHEP, BNL and JINR.

The Daya Bay experiment has been approved by the Chinese government, and tunnel construction will start in the spring of 2007. A fast run in 2009 using only one of the near sites and the mid hall is being considered. This offers an opportunity for studying background and systematic issues along with prompt physics results. The physics reach of this run is shown in Fig.5. The full deployment of the detectors for a three-year run is expected to be complete by the end of 2010.

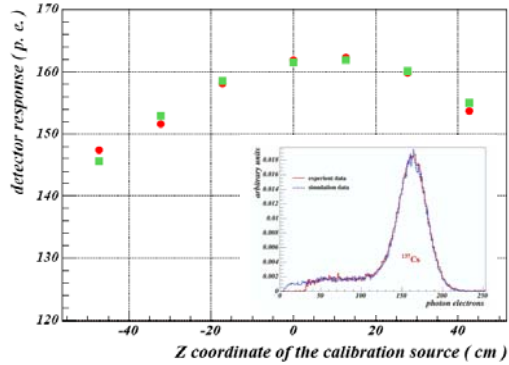

Fig. 6. Test results from the prototype module: The observed detector response compared to Monte Carlo simulations. Their excellent agreement shows that the optical reflector, light transport in the liquid scintillator, and the PMT response function etc. are all reasonably understood.

**References**

1. Y.F. Wang, Int. J. Mod. Phys. **A 20** (2005) 5244; K. Anderson *et al.,* White paper report on using reactors to search for a value of $\theta_{13}$, 2004, http://www.hep.anl.gov/minos/reactor13/white.html
2. L. Mikaelyan and V.V. Sinev, *Phys. Atom. Nucl.* **63**, 1002(2000), hep-ex/9908047.
3. Y.F. Wang, Nucl. Instr. And Meth. A 503(2003) 141; M.J. Chen et al., Nucl. Instr. And Meth. **A 562** (2006) 214.
4. M. Apollonio et al., Eur. Phys. J. **C27** (2003) 331.

---

[i] For the details of the Daya Bay experiment and a complete list of collaborators, see http://dayabay.ihep.ac.cn.